\documentclass[twocolumn,prb,aps,showpacs,superscriptaddress]{revtex4-1}
\usepackage{graphicx}
\newcommand{\hm}{h_{mw}}

\usepackage[usenames,dvipsnames]{pstricks}
\usepackage{pst-plot}
\usepackage{epsfig}
\usepackage{pst-grad} 
\usepackage{pst-plot} 
\usepackage{pstricks-add}
\usepackage{lmodern} 
\newcommand{\mycol}{1}
\newcommand{\beq}{\begin{equation}}
\newcommand{\eeq}{\end{equation}}
\usepackage{amsmath}
\usepackage[english,american]{babel}

\begin{document}

\title{Forbidden coherent transfer observed between two realizations of quasi-harmonic spin systems}

\author{S. Bertaina}
\email{sylvain.bertaina@im2np.fr}
\affiliation{Aix-Marseille Universit\'{e}, CNRS, IM2NP (UMR 7334), Marseille, France}

\author{G. Yue}
\affiliation{Department of Physics and The National High Magnetic Field Laboratory, Florida State University, Tallahassee, Florida 32310, USA}

\author{C-E Dutoit}
\affiliation{Aix-Marseille Universit\'{e}, CNRS, IM2NP (UMR 7334), Marseille, France}

\author{I. Chiorescu}
\email{ic@magnet.fsu.edu}
\affiliation{Department of Physics and The National High Magnetic Field Laboratory, Florida State University, Tallahassee, Florida 32310, USA}

\date{\today}%

\begin{abstract} The multi-level system $^{55}$Mn$^{2+}$ is used to generate
	two pseudo-harmonic level systems, as representations of the same electronic
	sextuplet at different nuclear spin projections. The systems are coupled using
	a forbidden nuclear transition induced by the crystalline anisotropy. We
	demonstrate Rabi oscillations between the two representations in conditions
	similar to two coupled pseudo-harmonic quantum oscillators. Rabi oscillations
	are performed at a detuned pumping frequency which matches energy difference
	between electro-nuclear states of different oscillators. We measure a coupling
	stronger than the decoherence rate, to indicate the possibility of fast
	information exchange between the systems.  
\end{abstract}

\pacs{03.67.-a 71.70.Ch 75.10.Dg 76.30Da}

\maketitle

\section{Introduction}

Recent advances in single spin measurements in gated nanostructures
\cite{Muhonen2014} and quantum dots \cite{Goryca_PRL2014} show that spin-based
materials have impact in quantum technologies. One example is constituted by
multi-level spin systems which have well defined spin Hamiltonians, sufficiently
large to be used as multi-qubit implementations \cite{Leuenberger2001} and small
enough to be studied by exact numerical methods. When diluted in non-magnetic
matrices, electronic spins attain large coherence times \cite{Nellutla2007,
	Bertaina2007,Bertaina_NatLett_2008} and present the possibility to be coupled
coherently to nuclear spins. In such implementations, magnetic ions (such as
rare-earth elements) play an essential role\cite{Probst2015,Bertaina2009a} with
demonstrated capability to coherently exchange information between electron
spins and nuclei \cite{Dutt2007,Wolfowicz_PRL2015} as well as optical
photons\cite{Clausen2011a}. In the current work, we are focusing on a \emph{3d}
element, Mn, which has a very low anisotropy and thus less stringent conditions
for the orientation of the external field, an important flexibility for on-chip
applications.

$^{55}$Mn$^{+2}$ ions diluted in a MgO matrix show electron spin resonance (ESR)
transitions with $\Delta S_z=1$ and $\Delta m_I=0$ ($S_z$ and $m_I$ are
projections of the electronic and nuclear $S=I=5/2$ moments respectively) at
fields separated by the hyperfine interaction in $2I+1$ well-defined groups.
However, off-diagonal couplings in the spin Hamiltonian, such as anisotropy and
transverse fields terms, can activate forbidden transitions $\Delta m_I \ne 0$
and/or $\Delta S_z\ne1$. The spectroscopy of forbidden transitions in
MgO:Mn$^{2+}$ is discussed in Refs.~[\onlinecite{Wolga1964}] and
[\onlinecite{Smith1968}] and give important information on their transition
probabilities. Large $\Delta m_I$ electro-nuclear mixture can be achieved by
making use of crystal and transverse fields \cite{Fataftah2016} as well.

The forbidden transitions reflect a coupling between different representations
of the same multi-$S_z$ system as detailed below. Although the transfer
probability between electro-nuclear states is low, the coherence properties are
robust and in addition it allows maneuvering the Hamiltonian in and out of the
forbidden (coupled) region. By using time-domain techniques, we can perform
multi-photon and/or detuned Rabi oscillations of the electronic spin states
\cite{Bertaina2009, Bertaina2011}.

Here we analyze the feasibility of combining high spin electronic and nuclear
systems to demonstrate coherent exchange of information between electro-nuclear
states.  The measurements are done by using a two-tone technique we have
recently developed\cite{Bertaina_prb2015}. We present theoretical and
experimental evidence of Rabi oscillations and of a driven strong coupling
regime between states belonging to different $m_I$. Such forbidden transitions
are essential to make the entire Hilbert space available for quantum information
manipulation and towards the use of long-lived nuclear spin states for storage
and retrieval of quantum information.

\section{Rabi oscillations of the electro-nuclear transition.}

The electro-nuclear Hamiltonian\cite{Bertaina2009} of the $^{55}$Mn$^{2+}$ ions is:
\begin{equation}
\label{eq:1}
\begin{split}
\mathcal{H}=&H_{CF}+\gamma\vec{H}_0\cdot\vec{S}-A\vec{S}\cdot\vec{I}+\gamma_N\vec{H}_0\cdot\vec{I}\\
&+\gamma\vec{h}_{mw}\cdot\vec{S}	\cos(2\pi ft).
\end{split}	
\end{equation}
The first term is the crystal field, the second is the static Zeeman
interaction, the third is the hyperfine interaction the fourth is the nuclear
Zeeman interaction and the last one is the dynamical Zeeman interaction caused
by the microwave field. $\gamma=g\mu_B/h$ is the gyromagnetic ratio ($g=2.0014$
the $g$-factor, $\mu_B$ Bohr's magneton and $h$ Planck's constant),
$\gamma_N=g_N\mu_N/h$ is the nuclear gyromagnetic ratio,   $S_{x,y,z}$ are the
spin projection operators, $\vec{S}$ is the total spin, $A = 244$~MHz is the
hyperfine constant, $h_{mw}$ and $f$ represent the microwave amplitude and
frequency respectively, and $\vec{H}_0$ is the static field ($\vec{H}_0\perp
\vec{h}_{mw}$). $H_{CF}=a/6[S_x^4+S_y^4+S_z^4-S(S+1)(3S^2+1)/5]$ with
$a=55.7$~MHz the anisotropy constant, represents the crystal field anisotropy
which generates a small anharmonicity of the otherwise equally spaced Zeeman
levels $S_z=-5/2...5/2$.

The model describing the multiphoton Rabi oscillations observed in MgO:Mn$^{2+}$
was reported in Ref.~[\onlinecite{Bertaina2009}]. However, the electron-nuclear
forbidden transitions were dropped off from the model since their probability
are  weak compared to the multiphoton electronic transitions. In this work, the
hyperfine term of $\mathcal{H}$ is no longer neglected, leading to a full
Hamiltonian $S\otimes I$ with a dimensionality of 36.

Let us consider a quantum system with 36 states $| S_z\otimes I_z\rangle$, $S_z$
and $I_z$=$\left\{-5/2, -3/2, -1/2, 1/2, 3/2, 5/2 \right\}$, irradiated by an
electromagnetic field. The spin Hamiltonian $\mathcal{H}$ can be rewritten as:
\begin{equation}
\label{eq:Ham2}
\mathcal{H}=\hat{E}+\hat{V}(t)=\sum_{S_z,I_z=-5/2}^{5/2} E_{S_z,I_z}|S_z\otimes I_z\rangle\langle
S_z\otimes I_z|+\hat{V}(t), 
\end{equation}
with $E_{Sz,I_z}$ the static energy levels,
$\hat{V}(t)=\frac{\gamma}{2}h_{mw}(\hat{S}_++\hat{S}_-)\cos\left(2\pi
ft\right)$, $S_+/S_-$ the raising/lowering operators. Since $H_0\gg \hm$, we use
the rotating wave approximation (RWA) to make Eq.~(\ref{eq:Ham2}) to be time
independent. We apply the unitary transformation $U(t)=\exp(-i2\pi f \hat{S}_z
t)$ to the Hamiltonian (\ref{eq:Ham2}) \cite{Hicke2007, Leuenberger2003}:%
\begin{equation} \label{eq:5} 
\mathcal{H}_{RWA}=U\mathcal{H}
U^\dag+i\hbar\frac{\partial U}{\partial t}U^\dag 
\end{equation}
and perform exact diagonalization of $\mathcal{H}_{RWA}$. Coherent motion of spin projection $S_z$ is
analyzed using time-dependent Schr\"odinger equation and its FFT can reveal
multiple Rabi frequencies and beatings (see below).

\begin{figure}
	\centering
	\includegraphics[width=\mycol\columnwidth]{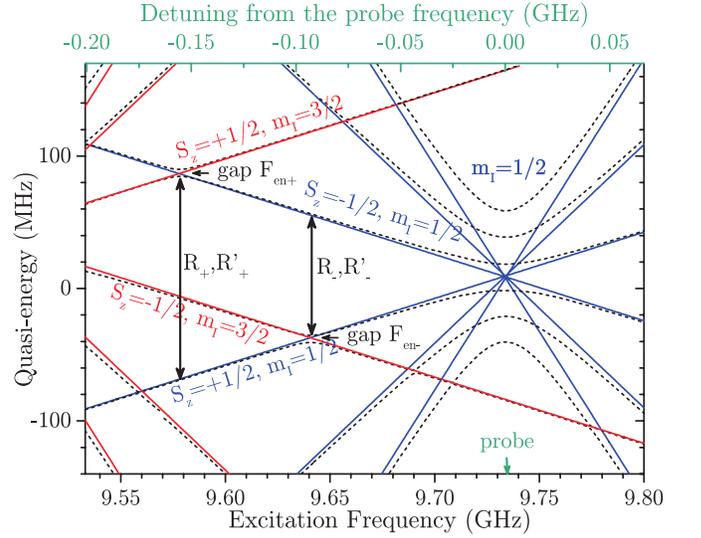}
	\caption{(color online) Quasi-energies of $H$ in the rotating wave approximation for low (lines)
		and high (dashed lines) microwave power for two sextuplets ${S_z}$: $m_I=1/2$
		(blue, right) and $3/2$ (red, left). The static field corresponds to the
		resonance frequency (shown by the green arrow) where the equally spaced levels
		collapse (for $h\rightarrow0$). Detuned Rabi oscillations, \emph{e.g.} $R_\pm$
		with location shown by the double-headed arrows, can be measured for any
		frequency in this range.} \label{fig:FIG_dressedstates}
\end{figure}
\begin{figure}
	\centering
	\includegraphics[width=\mycol\columnwidth]{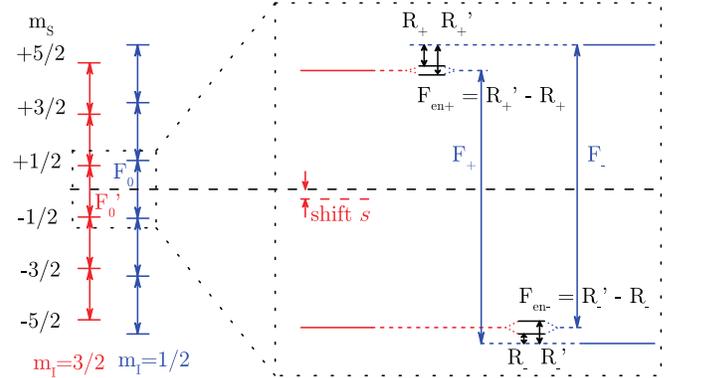}
	\caption{(color online) Sketch of $H$ eigen-states for $m_I=1/2$ and $3/2$. The central four
		states are magnified, showing the relationships $F_0=F_-+R_-=F_++R_+=F_0'+A$,
		as well as a splitting $F_{en\pm}$ of Rabi frequencies. The shift $s$ (see text) enforces
		$R_+\neq R_-$ and thus distinct double-headed arrows in
		Fig.~\ref{fig:FIG_dressedstates}. } \label{fig:fig_schema_en}
\end{figure}

Previous work \cite{Bertaina2009} shows that at exactly the ``compensation
angle" $\theta=\theta_c$ between $\vec{H}_0$ and the crystal axis $z$, the
anharmonic effect of the $H_{CF}$ term is compensated for and the $S_z$ levels
are equidistant. In the present study, we chose to work at this compensation
angle to reduce the level structure to simple pseudo-harmonic systems and put in
evidence their coupling. Note that the applied static field ensures a Zeeman
splitting of $\gamma H_0\approx f\sim10$~GHz, much stronger than all other
interactions of Eq.(\ref{eq:1}), effectively making $\vec{H}_0$ the quantization
axis.

In the RWA approximation, the resonance is shown in
Fig.~\ref{fig:FIG_dressedstates} by the probe arrow where all eigen-states
collapse (for $m_I=1/2$, the blue sextuplet). The two sextuplets are two
pseudo-harmonic oscillators as different realizations of the same set
${|S_z\rangle}$. The dashed lines indicate the effect of a large $h_{mw}$ on the
numerically computed dressed states. Note that Fig.~\ref{fig:FIG_dressedstates}
shows the levels in RWA while Fig.~\ref{fig:fig_schema_en} is a sketch showing
the laboratory frame picture. The assignment of spin projections $S_z$ in
Fig.~\ref{fig:FIG_dressedstates} correlates to the slopes of the quasi-energies
dependence on detuning. In Fig.~\ref{fig:fig_schema_en}, the two electro-nuclear
transitions appear in diagonal (see the levels connected by $F_+$and $F_-$
respectively), rather than as avoided level crossings as in
Fig.~\ref{fig:FIG_dressedstates}.

The nature and magnitude of the forbidden transition probabilities have been studied theoretically in this system\cite{Smith1968} and they follow (here $\theta=\theta_c=31^\circ$):
\beq
F_{en\pm}\propto 5\sin4\theta(ah_{mw}/f) [I(I+1)-m_I(m_I-1)].
\label{eq:2}
\eeq
Thus, the small anisotropy $a$ is the essential ingredient to the coupling
between the two oscillators, by enabling forbidden electronic transitions with
$\Delta S_z=\Delta m_I=1$. Notable is the dependence of $F_{en\pm}$ on microwave
field $h_{mw}$ which allows \emph{in-situ} control over the strength of coupling
between the two pseudo-harmonic oscillators. 

The experimental data discussed below, suggest that the dynamics of the group
of four states shown in Fig.~\ref{fig:fig_schema_en} can be driven
independently from the other levels. This leads to a description in terms of an
effective $4\times4$ RFA Hamiltonian:
\beq
\label{eq:3}
H_{RF}=\Delta s_z-As_zi_z-si_z+2F_{en}s_xi_x+\gamma h_{mw}s_x 
\eeq
where $F_{en}=F_{en\pm}$, $s_{x,z}$ and $i_{x,z}$ are the Pauli matrices for the
electronic and nuclear spin operator in this effective representation and
$\Delta=f-(F_0-A/2)$ is the detuning of the pump pulse by respect to $F_0-A/2$.
The coupled oscillations take place at $\Delta, A, s, F_{en}\gg \gamma h_{mw}$
in which case analytical diagonalization of $H_{RF}$ for $h_{mw}\sim0$ leads to
two pairs of levels:
\begin{align}
S_\pm^{(1)} & =A/4\pm\frac{1}{2}\sqrt{(\Delta+s)^2+F_{en}^2}\nonumber \\
S_\pm^{(2)} & =-A/4\pm\frac{1}{2} \sqrt{(\Delta-s)^2+F_{en}^2},       
\end{align}
with $F_{en}$ given by Eq.~\ref{eq:2}. The splitting of each pair at $\Delta=\pm s$ is
$F_{en}$, as expected. 

Different from the strong coupling regime in cavity QED
experiments~\cite{Chiorescu2010,Kubo2010,Schuster2010} is the fact that here
the Vaccum Rabi Splitting is observed under a sufficiently strong drive, since
$F_{en\pm}\propto h_{mw}$. Resonant photons of energy $hF_-$ (or $hF_+$) match
the difference between $|0,e\rangle$ and $|1,g\rangle$, where $0$ and $1$ label
the state of the pseudo-harmonic electronic spin system (operator $s_z$) and
$e,g$ label the nuclear state (operator $i_z$). The analogy could be further
extend by taking into consideration the multi-level structure of each $m_I$
subset, with $S_z=-5/2\dots5/2$. The $F_{en}$ transition allows the exchange of
information between subsets, to be followed by spin manipulation within a
subset\cite{Bertaina2009, Bertaina2011}.

\section{Experimental procedure}
Measurements were performed using a conventional Bruker Elexsys 680 pulse
spectrometer. The second frequency source is provided by the ELDOR bridge of
the spectrometer. The experiments are performed in a static field corresponding
to $F_0=9.734$~GHz, $F_-=9.641$~GHz and $F_+=9.586$~GHz. In fixed static field,
a first ESR pulse excites the system at any frequency detuning and a second
pulse reads the difference in level population at the main resonance. As
explained below, this method allows us to detune two representations of
multi-level systems until they are brought in resonance, to demonstrate the
strong coupling regime and a coherent transfer of information between the two
systems. The temperature was set to 50~K to have the relaxation time $T_1$ long
enough to perform the pulse sequence.

A first microwave pump pulse of frequency $f=F_\pm$ drives Rabi oscillations of
the Mn$^{2+}$ spins. In order to induce the coherent manipulation of an
electro-nuclear forbidden transition, we set the microwave power to the maximum
value available on the spectrometer. Because of the presence of a resonant
cavity the amplitude of the microwave field depends on the frequency and the
cavity transfer function. The $h_{mw}$ calibration has been done using the
following procedure: we measured at maximum microwave power the nutation frequency of a $S=1/2$
calibration standard (DPPH) by sweeping the microwave frequency $f$  and
the static field $H_0$ to keep them in resonant condition. Using the relation
between nutation frequency and microwave field: $g\mu_Bh_{mw}(f)/2= h \Omega_R(f)$
we found the microwave field amplitude as a function of the
microwave frequency used for the pumping pulse. In particular we found for
$f=F_0$, $h_{mw}\sim 20$~G, for $f=F_-$, $h_{mw}\sim 13$~G and for $f=F_+$,
$h_{mw}\sim 10$~G.

After a time longer than the Rabi decay time but much shorter than the
relaxation time $T_1$, a second  pulse ($\pi/2$ in 20~ns) at $f=F_0$ probes the
$S_z$ component, using the intensity of the Free Induction Decay (FID) signal.
The second pulse does not require a high power since it probes an allowed
transition\cite{Bertaina_prb2015} $m_s=-1/2 \rightarrow 1/2$. The sample is a
($2\times2\times1$)~mm$^3$ single crystal of MgO doped with Mn$^{2+}$ in a small
concentration of $\sim10^{-5}$; the crystal is oriented such that the allowed
transitions appear at the same field (see the ``compensation angle'' $\theta_c$
above).

\begin{figure} \centering
	
	\includegraphics[width=\mycol\columnwidth]{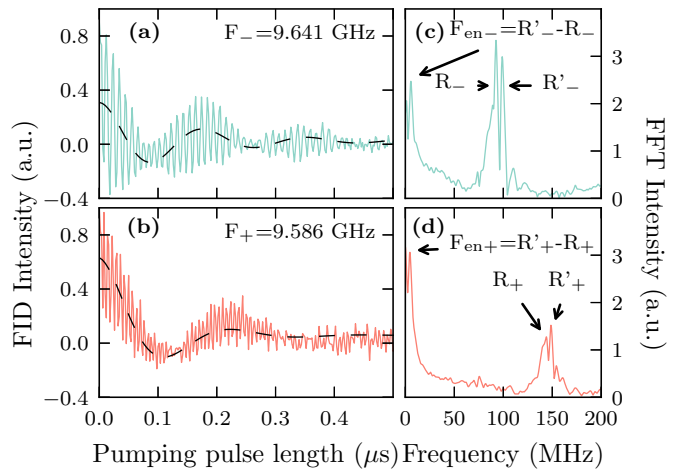} \caption{(color online)
		Rabi oscillations (left panel) for pump frequencies $F_{\pm}$ matching
		forbidden resonance conditions between two pseudo-harmonic level systems (see
		Fig.~\ref{fig:fig_schema_en}). The dashed line is a beating between two Rabi
		frequencies, demonstrating a coupling between the two systems stronger than
		the Rabi decay time. Corresponding FFT spectra are shown in the left panel,
		with visible strong coupling splittings $F_{en\pm}=R_\pm' - R_\pm$.}
	\label{fig:FIG_1Dtimedepedence} \end{figure}

For clarity, only the case of $m_I=3/2$ and $1/2$ is presented here. The
sextuplets are sketched in Fig.~\ref{fig:fig_schema_en} together with a
magnified representation of the main ESR transition, between states
$S_Z=\pm1/2$. The Zeeman splittings are $F_0=f$ and $F_0'=F_0-A$ for $m_I=1/2$
and $3/2$ respectively. The two transitions are shifted by an amount $s$ which
is evaluated by Drumheller\cite{PhysRev.133.A1099} as a second order
perturbation in $A$ and generates the vertical shift between sextuplets in
Fig.~\ref{fig:FIG_dressedstates}. The shift $s\approx33$~MHz is in good
agreement with the theoretical estimation\cite{PhysRev.133.A1099} of
$\sim44$~MHz. Consequently, forbidden couplings between sextuplets occur at
different frequencies $F_\pm=F_0-A/2\mp s$, allowing their individual
excitation by an adequate detuning of the drive frequency $f$. The frequency of
Rabi oscillations is shown by double headed arrows $R_{\pm}=F_0-F_\pm=A/2\pm
s$. As highly detuned Rabi oscillations, their frequency depends almost
linearly on the detuning from probe frequency $F_0$.

Moreover, if the electro-nuclear coupling between the two sextuplets is larger
than the decoherence rate, a splitting of the Rabi frequencies should be
observed ($F_{en\pm}=R_\pm'-R_\pm$) as illustrated in
Fig.~\ref{fig:fig_schema_en} and numerically calculated in
Fig.~\ref{fig:FIG_dressedstates}.

\section{Results and discussions}

\begin{figure} \centering \includegraphics[width=\mycol\columnwidth]{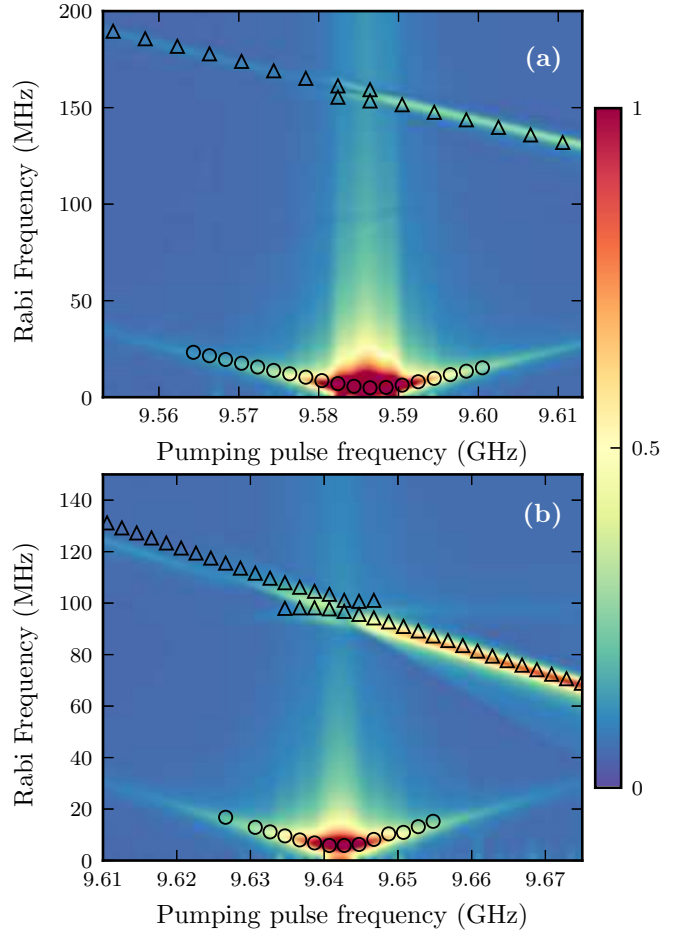}
	\caption{(color online) FFT spectra as a function of drive frequency around
		$F_+$ (a) and $F_-$ (b). The peaks location, intensity and width indicate Rabi
		oscillations frequency, amplitude and decay rate. The contour plot represents
		FFT of simulated Rabi oscillations (the color coded scale is in arbitrary
		units). The symbols are FFT peaks of measured Rabi oscillations: $_\triangle$ /
		$\circ$  correspond to the high / low frequency Rabi oscillations shown in 
		Fig.~\ref{fig:FIG_1Dtimedepedence}} \label{fig:FIG_zoom958} \end{figure}

\begin{figure} \centering \includegraphics[width=\mycol\columnwidth]{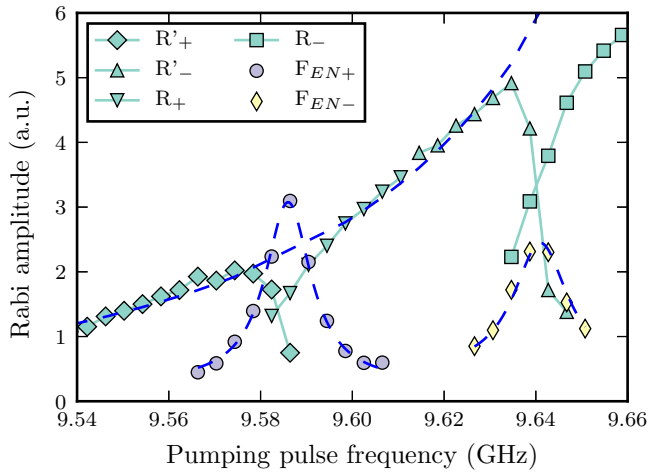}
	\caption{(color online) Amplitudes of Rabi oscillations (FFT peak intensities
		from Fig.~\ref{fig:FIG_zoom958}) as a function of drive frequency, fitted by a
		TLS model (\ref{eq:P}) (dotted and dashed lines). In the strong coupling
		region, $F_{en\pm}$ are very well described by the fit.}
	\label{fig:FIG_intensities} \end{figure}

Experimental results are presented in Fig.~\ref{fig:FIG_1Dtimedepedence} as
time-domain Rabi oscillations (left panel) and their Fast Fourier Transform
(FFT) spectra (right panel). One observes that the Rabi frequencies are
relatively large for a spin system, with $R_-\approx100$~MHz and
$R_+\approx150$~MHz, although they are in a detuned regime (the Rabi frequency
at resonance is $\approx34$~MHz). More importantly, the coupling between the
two electronic systems is sufficiently strong to overpass the Rabi decay rate
of $\Gamma_R\sim1/(250$~ns). This leads to normal mode splitting of the two
Rabi frequencies by an amount $F_{en\pm}=R_\pm'-R_\pm=4.5$~MHz and 5.7~MHz,
respectively. In the left panel of Fig.~\ref{fig:FIG_1Dtimedepedence}, the
dashed line shows the $F_{en+}$ and $F_{en-}$ beatings with a damping of
$\sim200$~ns and $\sim250$~ns respectively.

The two-tone technique allows the study of coupled Rabi oscillations for any
detuning, in the vicinity of $F_{\pm}$. FFT spectra are shown in
Fig.~\ref{fig:FIG_zoom958}, as a function of drive frequency around the
resonances $F_+$ (a) and $F_-$ (b). The symbols represent the values of
$R_{\pm}, R_{\pm}'$ (triangles) and $F_{en}$ (circles) as FFT peaks of measured
Rabi dynamics while the contour plot is calculated by exact diagonalization of
$H$ in the rotating frame. The simulations are resolving for the value of the
Rabi peaks and less for their intensities, color coded from blue to dark red
(arbitrary units) in Fig. 4. One notes how the Rabi oscillation accelerates with
the detuning to large values, close to 200~MHz. The low frequency beating
$F_{en}$ is equal to the splitting of the two Rabi frequencies, when the
resonance condition described above
(Figs.~\ref{fig:FIG_dressedstates},\ref{fig:fig_schema_en}) is met. Moreover,
one observes an increase of the beat frequency due to detuning, similar to the
coherent motion of a two-level system (TLS) driven out of resonance. Here,
$F_{en\pm}$ represent the coherent motion between two pseudo-harmonic
oscillators.

The height of a FFT peak in Fig.~\ref{fig:FIG_zoom958} represents the amplitude
of the corresponding Rabi oscillation. They are extracted and shown in
Fig.~\ref{fig:FIG_intensities} as a function of detuning away from $F_0$. While
Rabi frequency increase with detuning, the oscillation's amplitude decreases.
For a TLS, the amplitude $P$ of the Rabi frequency $F_R(\delta)$ as a function
of detuning $\delta$ is described by the well-known
relation\cite{Cohen-Tannoudji2006}: \beq
P=a_1\frac{F_R(0)^2}{F_R(0)^2+\delta^2}+a_0 \label{eq:P} \eeq with
$a_{i=0,1}=i$ in the Rabi model, corresponding to a full swing spin-up
$\leftrightarrow$ spin-down at resonance, while here they are fit parameters.
The detuning $\delta$ is defined as the difference between the pump frequency
$f$ and the resonance frequency of the considered TLS: $F_\pm$ for the
electro-nuclear transitions and $F_0$ for the spin transition.
	
The three situations are fitted very well by Eq.~\ref{eq:P}, as shown by dashed
curves in Fig.~\ref{fig:FIG_intensities}. The TLS model describes the $R_{\pm}$
and $R_{\pm}'$ data (dotted curve, $a_1\approx62$ and $a_0\approx0$) in regions
where the dynamics $R_{\pm}$ is not affected by the strong coupling of the two
level systems. The fitted half-width at half-maximum (HWHM) gives a Rabi
frequency at resonance of 31~MHz, close to a measured value of 34~MHz and in
agreement with the amount of power estimated in the cavity.

In the coupled region $f\sim F_{\pm}$, the extracted HWHM also gives very good
estimations of the beat frequencies: $F_{en+}=6.06$~MHz and $F_{en-}=7.2$~MHz.
The other fit parameters are $a_1=2.8$ and 2, $a_0=0.27$ and 0.44, for $F_+$
and $F_-$ respectively.

Since the TLS model (\ref{eq:P}) applies well to the forbidden transitions, the
dynamics shown in Fig.~\ref{fig:fig_schema_en} can be described by the
effective Hamiltonian given in Sect.~II. Following Eq.~\ref{eq:2}, the
couplings $F_{en\pm}$ should be equal while in our experiment they are slightly
different. This is due to the cavity resonance profile, centered around $\sim
F_0$ and detuned at $F_{\pm}$ with $F_+<F_-<F_0$. Consequently, the microwave
fields at $F_\pm$ follow $h_{mw+}<h_{mw-}$ which leads to $F_{en+}<F_{en-}$.
This case can be described by Hamiltonian~(\ref{eq:3}) by replacing the term
$2F_{en}s_xi_x$ with $F_{en-}(s_+i_-+s_-i_+)/2+F_{en+}(s_+i_++s_-i_-)/2$.

In this case, the condition to tune the systems into the strong coupling
regime, is to have $h_{mw\pm}$ sufficiently large such that
$F_{en\pm}>\Gamma_R$ (defined above), condition indeed fulfilled in our
experiment. The decay rate of individual (or coupled) oscillations is mostly
due to the inhomogeneity of the microwave field
\cite{DeRaedt2012,Baibekov2011}. Volume integration over the entire spin
population, causes a fast $\Gamma_R$ although the echo-detected coherence times
are 1-2 orders of magnitude larger. To detect faster or larger coupled
forbidden oscillations in the system presented here, one can in principle
utilize setups providing larger power or sensitivity allowing the study of
samples with smaller volume (and thus smaller $h_{mw}$ inhomogeneity) or lower
Mn doping concentrations (and thus lower level of long-range dipolar
interactions)..

\section{Conclusion} We show coherent transfer of state population between two
equidistant level systems, $|S_z,m_I=1/2\rangle$ and $|S_z,m_I=3/2\rangle$ by
using forbidden nuclear transitions with $\Delta m_I=1$. The coupling between
systems is tunable and is stronger than the decay rate, leading to an
observable splitting of the Rabi mode and a state transfer faster than the
decay time. The results open the way of combining the electronic and long-lived
nuclear degrees of freedom in this multi-level system.

\section*{Acknowledgements} This work was supported by NSF Grant No.
DMR-1206267, CNRS-PICS CoDyLow and CNRS's research federation RENARD (FR3443)
for EPR facilities. The NHMFL is supported by the Cooperative Agreement Grant
No. DMR-1157490 and the State of Florida.


\end{document}